\documentstyle[12pt]{article}
\begin{document}
\footskip50pt
\parindent=12pt
\parskip=.4cm
\input amssym.def
\input amssym
\def\tC{\tilde{C}}
\def\bR{{\Bbb R}}
\def\a{\"{a}}
\def\o{\"{o}}
\def\u{\"{u}}
\def\ep{\epsilon}
\def\l{\langle}
\def\r{\rangle}
\def\g{g^{\mu \nu}}
\def\d{{\rm d}}
\def\D{{\cal D}}
\def\L{{\cal L}}
\def\S{{\cal S}}
\def\J{{\cal J}}
\def\T{{\cal T}}
\def\X{{\cal X}}
\def\Y{{\cal Y}}
\def\ome{{\cal A}}
\def\Ga{\Gamma}
\def\pr{\partial}
\def\O{{\cal O}}
\def\R{{\cal R}}
\def\nab{\nabla}
\def\I{{\cal I}}
\def\J{{\cal J}}
\def\E{{\cal E}}
\def\hU{\hat{U}}
\def\hV{\hat{V}}
\def\hh{{1\over 2}}
\def\half{{\textstyle{ 1 \over 2}}}
\def\quar{{\textstyle{ 1 \over 4}}}
\def\ts{\textstyle}
\def\A{{\cal A}}
\def\B{{\cal B}}
\def\C{{\cal C}}
\def\de{\delta}
\def\si{\sigma}
\def\ga{\gamma}
\def\la{\lambda}
\def\ka{\kappa}
\def\be{\beta}
\def\topcirc{\mathaccent"7017}
\def\oT{{\topcirc T}}
\def\oG{{\topcirc G}}
\def\oD{{\topcirc \D}}
\def\oDel{{\topcirc \Delta}}

\begin{flushright}
DAMTP/97-81, gr-qc/9708040
\end{flushright}

\begin{center}
{\Large \bf Conformally Covariant Differential Operators:}\break
{\Large \bf Symmetric Tensor Fields} 
\end{center}

\vspace{1.5cm}

\begin{center}
Johanna Erdmenger 

Institut f\u r Theoretische Physik, 
Universit\a t Leipzig, Augustusplatz 10/11, \newline
D - 04109 Leipzig, Germany. \
e-mail: erd@tph204.physik.uni-leipzig.de 

and 

Hugh Osborn

Department of Applied Mathematics and Theoretical Physics, University
of Cambridge, Silver Street, Cambridge CB3 9EW, England. \ 
e-mail: ho@damtp.cam.ac.uk 

\end{center}

\vspace{1.5cm}

\begin{center} \bf Abstract \end{center}

We extend previous work on conformally covariant differential operators
to consider the case of second order operators acting on symmetric traceless
tensor fields. The corresponding flat space Green function is explicitly
constructed and shown to be in accord with the requirements of conformal
invariance.

\rm \normalsize

\begin{flushleft}
\end{flushleft}

\newpage

Conformal differential operators $\Delta$ are covariant differential 
operators acting on tensor fields, or more generally sections of some vector
bundle, over a curved manifold with metric $g_{\mu\nu}$ which also  transform
covariantly under local Weyl rescalings of the metric
\begin{equation}
\de_\si g^{\mu\nu} = 2 \si \, g^{\mu\nu} \, ,
\label{one}
\end{equation}
so that $\de_\si \Delta= r \Delta \si + (s-r)\si \Delta$ for some $r$ if 
$\Delta$ is $s\,$th order. Such operators are generalisations of the well
known operator $-\nab^2 + \frac{1}{6}R$ acting on scalars in four dimensions and
have been classified by Branson \cite{Branson}. 
Except for special values of the dimension $d$
such operators exist for general tensor fields belonging to representations
of the tangent space group $O(d)$, or $O(d-1,1)$, or their spinor counterparts.
The Greens functions associated with such conformal differential operators also
transform covariantly under local rescalings of the metric and they may have
a role in constructing forms for the quantum field theory
effective action on curved manifolds.

Recently one of us discussed conformal differential operators and their
associated Green functions from the point of view of the reduction to flat
space \cite{Joh} (this paper is subsequently referred to as I). 
In this case the form of the flat space Green function is unique up to an
overall constant due to the restrictions imposed by the flat space conformal
group $O(d+1,1)$ or $O(d,2)$ \cite{OP,EO}. In I the general analysis was applied to
conformal differential operators acting on totally antisymmetric $k$-index
tensor fields, or $k$-forms, and also for 4-index tensor fields with the
symmetries of the Weyl tensor, in both cases for arbitrary dimension $d$ when
the general results were explicitly verified.

In this follow up we extend the discussion to totally symmetric, traceless
$p$-index tensor fields and again find a result for the flat space Green function
which is in accord with gene\-ral theory, although the combinatorics are more
involved in this case. The corresponding conformal differential operator was
apparently first constructed by  W\"unsch \cite{Wuensch} and also found as
part of his general theory by Branson \cite{Branson}. For $p=2$ a particular
conformally covariant differential operator was found by Gusynin and
Roman'kov \cite{Gusynin} (the general case involves a term proportional to
the Weyl tensor with an arbitrary coefficient), for $d=4$ see also \cite{Deser}.
The case of general $p$ has also been discussed more
recently from a rather different point of view
by O'Raifeartaigh {\it et al} \cite{Raifertaigh}, the results agree with ours
when $d=4$.  For completeness we here follow, for arbitrary $d$,  
the general method of \cite{Raifertaigh} which determines the
conformal differential operator $\Delta^S$ by first constructing a Weyl
invariant quadratic action $S^S(g,\omega)$ for the symmetric traceless tensor 
field $\omega_{\mu_1\dots \mu_p}$. With a convenient overall normalisation a
general expression with manifest coordinate invariance which is second order
in covariant derivatives has the form
\begin{equation}
S_0[g,\omega] \, = \, \frac{1}{2p!} \int \! \d^d x \,
\sqrt g \, \Big[ \, \nabla^\lambda \omega^{\mu_1\dots \mu_p}
\nabla_\lambda \omega_{\mu_1\dots \mu_p} \, + \, a \, \nabla_\rho
\omega^{\mu_1\dots \mu_{p-1}\rho} \nabla^\lambda
 \omega_{\mu_1\dots \mu_{p-1}\lambda} \Big] \, ,
\label{S0}
\end{equation}
for $a$ an arbitrary parameter.
Since from (\ref{one})
$\de_\si \sqrt g = - d \si \sqrt g$ it is easy to see that this is invariant
under constant Weyl rescalings if
\begin{equation}
\de_\si \omega_{\mu_1\dots \mu_p} = \half (d-2p-2) \si \, \omega_{\mu_1\dots \mu_p}
\, .
\label{two}
\end{equation}
In general an action which is invariant under rigid scale transformations
has a variation under local $\si(x)$ linear in derivatives of the form
\begin{equation}
\de_\si S_0[g,\omega] = \int \! \d^d x \, \sqrt g \, \pr_\lambda \si J^\lambda \, .
\end{equation}
For the action given in (\ref{S0}) the derivatives and Christoffel connections 
generate an explicit expression for $J^\lambda$,
\begin{eqnarray}
p!\, J^\lambda\!\! &=& \!\!  
\half(d-2) \nabla^\lambda \omega^{\mu_1\dots \mu_p}
\, \omega_{\mu_1\dots \mu_p} + p \,  \nab^\rho
\omega^{\mu_1\dots \mu_{p-1}\lambda} \, \omega_{\mu_1\dots \mu_{p-1}\rho}
\nonumber \\
&& \qquad{}
- \Big ( p + \half (d+2p-2)a\Big ) \,  
\omega^{\mu_1\dots \mu_{p-1}\lambda} \, 
\nabla^\rho \omega_{\mu_1\dots \mu_{p-1}\rho}  \, .
\label{var0}
\end{eqnarray}
In order to achieve a Weyl invariant action it is essential to be able to
re-express the variation in terms of second derivatives of $\si$. To achieve
this it is necessary that
\begin{equation}
J^\lambda = \nab_\rho \J^{\lambda\rho} \quad \Rightarrow \quad
\de_\si S_0[g,\omega] = - \int \! \d^d x \, \sqrt g \, \nab_\rho \pr_\lambda \si 
\J^{\lambda\rho} \, ,
\label{var1}
\end{equation}
where clearly we may assume $\J^{\lambda\rho} = \J^{\rho\lambda}$. In this
case the variation in (\ref{var1})
may be cancelled by an additional curvature dependent action. From (\ref{var0})
it is easy to see that the result in (\ref{var1}) is possible only if
\begin{equation}
a = - \frac{4p}{d+2p-2} \, ,
\label{a}
\end{equation}
and then
\begin{equation}
p!\, \J^{\lambda\rho}  = 
\quar(d-2) \, \de^{\lambda\rho} \, 
\omega^{\mu_1\dots \mu_p}\omega_{\mu_1\dots \mu_p}
+ p \, 
\omega^{\mu_1\dots \mu_{p-1}\lambda} \omega_{\mu_1\dots \mu_{p-1}}{}^{\! \rho} 
\, .
\label{var}
\end{equation}
To exhibit the required curvature dependent terms it is convenient to define
in terms of the Ricci tensor $R_{\mu\nu}$ and scalar curvature $R$
\begin{equation}
J=  \frac{1}{2(d-1)} R \, , \qquad K_{\mu \nu} = \frac{1}{(d-2)} 
\left( R_{\mu \nu} - J g_{\mu \nu} \right) \, ,
\label{defJK}
\end{equation}
since these transform under local Weyl rescalings as in (\ref{one}) according to
\begin{equation} \label{}
\delta_\si J = 2 \si J + \nabla^2 \sigma \, , \qquad \delta_\si K_{\mu\nu} =
\nabla_\mu \nabla_\nu \sigma \, .
\end{equation}
It is then evident that the action 
\begin{equation}
S_1[g,\omega] \, = \, \frac{1}{2p!} \int \! \d^d x \,
\sqrt g \, \Big[ \half(d-2)J \, \omega^{\mu_1\dots \mu_p}
\omega_{\mu_1\dots \mu_p} \, + \, 2p \, K_{\rho\lambda}\, 
\omega^{\mu_1\dots \mu_{p-1}\rho} 
\omega_{\mu_1\dots \mu_{p-1}}{}^{\!\lambda} \Big] \, .
\label{S1}
\end{equation}
has a variation which exactly compensates that given by (\ref{var1}) and
(\ref{var}). We may also add a contribution which is separately invariant under
local Weyl rescalings
\begin{equation} 
S_2[g,\omega] \, = \, \frac{A}{2p!} \int \! \d^d x \, \sqrt g \, 
C_{\lambda\ep\rho\eta} \omega^{\mu_1\dots \mu_{p-2}\lambda\rho}
\omega_{\mu_1\dots \mu_{p-2}}{}^{\! \ep\eta} \, ,
\label{S2}
\end{equation}
which depends on the Weyl tensor which may be defined by
\begin{equation}
C_{\alpha \beta \gamma \delta} = R_{\alpha \beta \gamma \delta}
- {2} \left( g_{\alpha [ \gamma} K_{\delta ] \beta} -
g_{\beta [ \gamma} K_{\delta ] \alpha} \right)  \, , 
\label{Weyl}
\end{equation}
The invariance of (\ref{S2}) follows simply since $\de_\si C_{\lambda\ep\rho\eta}
= - 2\si \, C_{\lambda\ep\rho\eta}$.

Hence the combined action given by (\ref{S0},\ref{S1},\ref{S2}), with (\ref{a}),
\begin{equation}
S^S[g,\omega] = S_0[g,\omega]+ S_1[g,\omega] + S_2[g,\omega]
= \, \frac{1}{2p!} \int \! \d^d x \,
\sqrt g \, \omega^{\mu_1\dots \mu_p} (\Delta^S \omega)_{\mu_1\dots \mu_{p}}
\label{S}
\end{equation}
defines a conformally covariant
differential operator $\Delta^S$ on symmetric traceless
tensors, depending on a single parameter $A$, such that\footnote{The result
for the curvature dependent terms arising from (\ref{S1}) differs from that in
the papers of Branson \cite{Branson} but this seems to arise from a simple
arithmetic error.}
\begin{equation}
\delta_\si \Delta^S = \half (d+2p+2)\si \, \Delta^S - \half (d+2p-2)\,
\Delta^S \si \, .
\end{equation}

The corresponding Green function is defined in general by
\begin{equation}
{\ts\sqrt{g(x)}}  \Big( \Delta_x^{S} G^{S}
\Big){}_{\mu_1 \cdots \mu_p,}{}^{\! \nu_1 \cdots \nu_p} (x,y) =
\E^S{}_{\!\!\mu_1 \cdots \mu_p,}{}^{\!\nu_1\cdots \nu_p}\, \delta^d(x-y) \, ,
\label{green}
\end{equation}
where $\E^S$ is the projector onto totally symmetric  traceless $p$-index tensors.
This may be given explicitly by
\begin{eqnarray}
\E^S{}_{\!\! \mu_1 \dots \mu_p,}{}^{\nu_1 \dots \nu_p} \!\! &=& \!\!
\delta_{(\mu_1}{}^{\! (\nu_1} \dots \delta_{\mu_p)}{}^{\! \nu_p)}
\nonumber \\
&& \!\! {}+ \sum_{r=1}^{[\frac{1}{2}p]} \, \lambda_r \,
g_{(\mu_1\mu_2}\dots g_{\mu_{2r-1}\mu_{2r}} \,
g^{(\nu_1\nu_2}\dots g^{\nu_{2r-1}\nu_{2r}} \, \delta_{\mu_{2r+1}}
{}^{\! \nu_{2r+1}}\dots  \delta_{\mu_p)}{}^{\! \nu_p)} \, , \nonumber \\
\lambda_r  \!\! &=& \!\! (-1)^r \, \frac{ p! }
{2^r r!(p-2r)! \prod_{s=1}^{r} (d+2p-2-2s)} \, ,
\end{eqnarray}
for $[\frac{1}{2}p]$ the  integer part of $\frac{1}{2}p$. Under Weyl rescalings
this transforms as
\begin{eqnarray}
\de_\si G^S{}_{\!\!\mu_1 \cdots \mu_p,}{}^{\!\nu_1\cdots \nu_p}(x,y)\!\!& =&\!\!
\half (d+2p-2)\, \si (x) \, 
G^S{}_{\!\!\mu_1 \cdots \mu_p,}{}^{\!\nu_1\cdots \nu_p}(x,y) \nonumber \\
&& {} + \half (d-2p-2) \, \si(y) \,
G^S{}_{\!\!\mu_1 \cdots \mu_p,}{}^{\!\nu_1\cdots \nu_p}(x,y) \, .
\end{eqnarray}

We here determine the flat space form for $G^S$ following similar procedures
to I. In the flat space limit, $g_{\mu\nu}\to \de_{\mu\nu}$, and we may 
identify up and down indices. Explicitly $\Delta^S \to \oDel^S$ which is given
by
\begin{eqnarray}
(\oDel^S \omega )_{\mu_1 \cdots \mu_p} \!\!& =&\!\! -\pr^2
\omega_{\mu_1 \cdots \mu_p} + \frac{4p}{d+2p-2} \, \pr_\lambda \pr_{(\mu_1}
\omega_{\mu_2\cdots \mu_{p-1})\lambda} \nonumber \\
&& - \frac{4p(p-1)}{(d+2p-2)(d+2p-4)} \, \pr_\rho \pr_\lambda \,
\de_{(\mu_1\mu_2} \omega_{\mu_3 \dots \mu_p)\lambda\rho} \, ,
\end{eqnarray}
where the last term serves to ensure that the r.h.s. is traceless. Finding
the flat space Green function,
\begin{equation}
G^{S}{}_{\!\!\mu_1 \cdots \mu_p,\nu_1 \cdots \nu_p} (x,y) \Big |_{g=\delta}
= {\topcirc G}^S {}_{\!\!\mu_1 \cdots \mu_p,\nu_1 \cdots \nu_p} (x- y) \, ,
\label{G0}
\end{equation}
is equivalent to solving
\begin{equation}
(\oDel^S \omega )_{\mu_1 \cdots \mu_p} = \phi_{\mu_1 \cdots \mu_p} \, ,
\end{equation}
for arbitrary $\phi$ and to this end we may write the Fourier transform as
\begin{equation}
{\widetilde \omega}_{\mu_1 \cdots \mu_p}
 = a_0 \, \frac{1}{k^2} {\widetilde \phi}_{\mu_1 \cdots \mu_p} +
\sum_{r=1}^p a_r \, \frac{2^r}{k^{2(1+r)}} k_{(\mu_1} \dots k_{\mu_r}
{\widetilde \phi}_{\mu_{r+1} \dots \mu_p)\rho_1 \dots \rho_r}
k_{\rho_1} \dots k_{\rho_r}- \mbox{traces}(\mu_1\dots \mu_p)\, .
\label{e1}
\end{equation}
The coefficients $a_r$ are then determined by requiring
\begin{equation}
k^2 {\widetilde \omega}_{\mu_1 \cdots \mu_p} - \frac{4p}{d+2p-2} \, 
k _{(\mu_1}  {\widetilde \omega}_{\mu_2\cdots \mu_{p-1})\lambda} k_\lambda
- \mbox{traces}(\mu_1\dots \mu_p) = {\widetilde \phi}_{\mu_1 \cdots \mu_p} \, .
\label{e2}
\end{equation} 

To analyse (\ref{e1}) and (\ref{e2}) we first consider a symmetric traceless
$(p{-r})$-index tensor $\psi_{\nu_1 \dots \nu_{p-r}}$ and obtain
\begin{eqnarray}
\lefteqn{
\Big [ k_{(\mu_1} \dots k_{\mu_r} \psi_{\mu_{r+1} \dots \mu_{p-1} \lambda)}
- \mbox{traces}(\mu_1\dots \mu_{p-1} \lambda) \Big ]} \nonumber \\
&=& \!\! r \, k_{(\mu_1} \dots k_{\mu_{r-1}}  \psi_{\mu_r \dots \mu_{p-1})}
\, k_\lambda + (p-r)\, k_{(\mu_1} \dots k_{\mu_r} \psi_{\mu_{r+1} \dots \mu_{p-1})
\lambda } \nonumber \\
&&\!\!\!\!\! {} - \frac{r(r-1)}{d+2p-4} \, k^2 \, \de_{\lambda(\mu_1} 
k_{\mu_2} \dots k_{\mu_{r-1}} \psi_{\mu_r \dots \mu_{p-1})}
- \frac{2r(p-r)}{d+2p-4} \, \de_{\lambda(\mu_1}k_{\mu_2} \dots k_{\mu_r}
\psi_{\mu_{r+1} \dots \mu_{p-1})\rho}\, k_\rho \nonumber \\
&&\!\!\!\!\!{} - \mbox{traces}(\mu_1\dots \mu_{p-1}) \, .
\end{eqnarray}
Hence
\begin{eqnarray}
\lefteqn{
\Big [ k_{(\mu_1} \dots k_{\mu_r} \psi_{\mu_{r+1} \dots \mu_{p-1} \lambda)}
- \mbox{traces}(\mu_1\dots \mu_{p-1} \lambda) \Big ] k_\lambda} \nonumber \\
&=& \!\! \frac{r}{p}\, \frac{d+2p-r-3}{d+2p-4} \, k^2 \, 
k_{(\mu_1} \dots k_{\mu_{r-1}} \psi_{\mu_r \dots \mu_{p-1})} \nonumber \\
&&\!\!\!\!\!{} + \frac{p-r}{p}\, \frac{d+2p-2r-4}{d+2p-4} \,
k_{(\mu_1} \dots k_{\mu_r} \psi_{\mu_{r+1} \dots \mu_{p-1})\rho}\, k_\rho 
\nonumber \\
&&\!\!\!\!\!{} - \mbox{traces}(\mu_1\dots \mu_{p-1}) \, .
\end{eqnarray}
Using this to calculate the result of inserting (\ref{e1}) into (\ref{e2})
we get
\begin{equation}
a_0 =1
\label{a0}
\end{equation}
and
\begin{equation}
\left ( 1 - \frac{2p}{d+2p-2} \, \frac{r}{p}\, \frac{d+2p-r-3}{d+2p-4}
\right ) a_r = \frac{2p}{d+2p-2} \,
\frac{p-r}{p}\, \frac{d+2p-2r-4}{d+2p-4} \, a_{r-1} \, ,
\label{ar}
\end{equation}
which simplifies to
\begin{equation}
a_r = \frac{p-r+1}{\half d+p-r-2} \, a_{r-1} \, .
\end{equation}
It is then straightforward, with (\ref{a0}), to find
\begin{equation}
a_r = \prod_{j=1}^r \frac{p-j+1} {\half d+p-j-2} \, .
\end{equation}
With these results for $a_r$ the Fourier transform of ${\topcirc G}^S$, as given
by (\ref{G0}), becomes
\begin{eqnarray}
\lefteqn{
\!\!\!{\widetilde {{\topcirc G}{}^S}{}_{\!\! \mu_1 \dots \mu_p,\nu_1 \dots \nu_p}(k)
= \E^S{}_{\!\! \mu_1 \dots \mu_p,\nu_1 \dots \nu_p}\, \frac{1}{k^2}}} \nonumber \\
&& \!\!\!\!\!\!\!\!\! {}+ \sum_{r=1}^p a_r \,
\E^S{}_{\!\! \mu_1 \dots \mu_p,\ep_1 \dots \ep_r \lambda_{r+1} \dots
\lambda_p} \E^S{}_{\!\! \eta_1 \dots \eta_r \lambda_{r+1} \dots \lambda_p,
\nu_1 \dots \nu_p}
\, \frac{2^r}{k^{2(1+r)}} k_{\ep_1} \dots k_{\ep_r} k_{\eta_1} \dots k_{\eta_r}
\, .
\label{Gt}
\end{eqnarray}

The inversion of the Fourier transform in (\ref{Gt}) may be found with the aid
of
\begin{eqnarray}
\lefteqn{\!\!\frac{1}{(2\pi)^d} \int \! \d^d k \, e^{- ik{\cdot x}} \,
\frac{1}{k^{2(1+r)}}
k_{\alpha_1}\dots k_{\alpha_{2r}}}
\nonumber \\
\!\!\! &=&\!\!\!
\frac{\Gamma(\half d - 1)}{4\pi^{\frac{1}{2}d}(x^2)^{\frac{1}{2}d
-1}} \, \frac{(2r)!}{4^r r!} \sum_{s=0}^r \frac{(-4)^s (\half d-1)_s}
{(r-s)!(2s)!} \, \frac{1}{x^{2s}} x_{(\alpha_1}\dots x_{\alpha_{2s}}
\delta_{\alpha_{2s+1} \alpha_{2s+2}} \dots \delta_{\alpha_{2r-1} \alpha_{2r})} \, ,
\end{eqnarray}
for $(y)_s = \Gamma(y+s)/\Gamma(y)$, and
\begin{eqnarray}
\lefteqn{
\E^S{}_{\!\! \mu_1 \dots \mu_p,\alpha_1 \dots \alpha_r \lambda_{r+1} \dots
\lambda_p}
\E^S{}_{\!\! \alpha_{r+1} \dots \alpha_{2r} \lambda_{r+1} \dots \lambda_p,
\nu_1 \dots \nu_p}
x_{(\alpha_1}\dots x_{\alpha_{2s}}
\delta_{\alpha_{2s+1} \alpha_{2s+2}} \dots \delta_{\alpha_{2r-1} \alpha_{2r})}}
\nonumber \\
&=& \!\! 2^{r-s} \, \frac{(2s)!}{(2r)!} \, \bigg( \frac{r!}{s!}\bigg)^{\! 2}
\E^S{}_{\!\! \mu_1 \dots \mu_p,\ep_1 \dots \ep_s \lambda_{s+1} \dots
\lambda_p} \E^S{}_{\!\! \eta_1 \dots \eta_s \lambda_{s+1} \dots \lambda_p,
\nu_1 \dots \nu_p}
x_{\ep_1} \dots x_{\ep_s} x_{\eta_1} \dots x_{\eta_s} \, .
\end{eqnarray}
We therefore find
\begin{eqnarray}
\lefteqn{
\!\!\!\!\!\! {\topcirc G}^S{}_{\!\! \mu_1 \dots \mu_p,\nu_1 \dots \nu_p}(x)  =
\frac{\Gamma(\half d - 1)}{4\pi^{\frac{1}{2}d}(x^2)^{\frac{1}{2}d-1}}
\bigg \{ b_0 \, \E^S{}_{\!\! \mu_1 \dots \mu_p,\nu_1 \dots \nu_p}} \nonumber \\
&& \!\!\!\!\!
 {}+ \sum_{s=1}^p b_s \,
\E^S{}_{\!\! \mu_1 \dots \mu_p,\ep_1 \dots \ep_s \lambda_{s+1} \dots
\lambda_p} \E^S{}_{\!\! \eta_1 \dots \eta_s \lambda_{s+1} \dots \lambda_p,
\nu_1 \dots \nu_p}
\frac{(-2)^s}{x^{2s}} x_{\ep_1} \dots x_{\ep_s} x_{\eta_1} \dots x_{\eta_s}
 \bigg \} \, ,
\label{Gb}
\end{eqnarray}
where
\begin{equation}
b_0 = 1+ \sum_{r=1}^{p} a_r  \, , \qquad
b_s = (\half d-1)_s  \frac{1}{s!} \sum_{r=s}^p {r \choose s}\, a_r \, , \ s\ge 1
 \, .
\label{b}
\end{equation}

To calculate $b_s$ we may use induction on $p$. From (\ref{a0},\ref{ar}) it is
easy to see that
\begin{equation}
a_r^{(p)} = \frac{p}{\half d+p-3}\, a_{r-1}^{(p-1)}  \, , \quad
r=1, \dots p \, .
\end{equation}
Since
\begin{equation}
{r \choose s} = {{r-1} \choose {s-1}} + {{r-1} \choose s}
\end{equation}
we find from (\ref{b})
\begin{equation}
b_0^{(p)} = 1 + \frac{p}{\half d+p-3}\, b_0^{(p-1)} \, , \quad
b_s^{(p)} = \frac{p}{s} \,  \frac{\half d+s-2}{\half d+p-3}\, b_{s-1}^{(p-1)} +
\frac{p}{\half d+p-3}\, b_s^{(p-1)} \, , \ s=1,2 \dots .
\end{equation}
It is then easy to verify the general result for any $p$
\begin{equation}
b_s = {p\choose s} \, \frac {\half d+p-2}{\half d - 2} \,.
\label{bres}
\end{equation}
Applying (\ref{bres}) in (\ref{Gb}) with the standard binomial theorem gives
\begin{equation}
{\topcirc G}^S{}_{\!\! \mu_1 \dots \mu_p,\nu_1 \dots \nu_p}(x)  =
\frac{\Gamma(\half d - 1)}{4\pi^{\frac{1}{2}d}(x^2)^{\frac{1}{2}d-1}}
\, \frac {d+2p-4}{d-4} \,
\I^S{}_{\!\! \mu_1 \dots \mu_p,\nu_1 \dots \nu_p}(x) \, ,
\label{GI}
\end{equation}
where
\begin{equation}
\I^S{}_{\!\! \mu_1 \dots \mu_p,\nu_1 \dots \nu_p}(x) = 
\E^S{}_{\!\! \mu_1 \dots \mu_p,\ep_1 \dots \ep_p} I_{\ep_1 \nu_1}(x) \dots
I_{\ep_p \dots \nu_p}(x) \, ,
\label{IS}
\end{equation}
for
\begin{equation}
I_{\ep\nu}(x) = \de_{\ep\nu} -  \frac{2}{x^2}\, x_\ep x_\nu \, .
\end{equation}
$I_{\ep\nu}(x)$ is the inversion tensor so that $
\I^S{}_{\!\! \mu_1 \dots \mu_p,\nu_1 \dots \nu_p}(x)$ as given by (\ref{IS})
is the inversion tensor for totally symmetric traceless $p$-index tensor fields
and (\ref{GI}) is exactly of the form expected as a consequence of applying
flat space conformal invariance in this case. Except when $p=0$ the Green
function does not exist for $d=4$ reflecting the fact that $d=4$ is the
critical dimension for $\Delta^S$.

To understand the role of the critical dimension $d=4$ we may introduce a
linear differential operator $\D$ acting on symmetric traceless tensors, with
index $p\ge 1$, which is defined by
\begin{eqnarray}
(\D \omega)_{\mu_1 \dots \mu_p \lambda} \!\! &=& \!\!
\nab_\lambda \omega_{\mu_1 \dots \mu_p} - \nab_{(\mu_1} \omega_{\mu_2 \dots 
\mu_p)\lambda} \nonumber \\
&& \!\! {} - \frac{p-1}{d+p-3} \Big ( g_{\lambda(\mu_1}
\nab^\rho \omega_{\mu_2 \dots \mu_p)\rho } - g_{(\mu_1\mu_2}
\nab^\rho \omega_{\mu_3 \dots \mu_p)\lambda\rho} \Big ) \, .
\label{defD}
\end{eqnarray}
This satisfies the traceless conditions
\begin{equation}
g^{\mu_1\mu_2} (\D \omega)_{\mu_1 \dots \mu_p \lambda} =
g^{\lambda \mu_1} (\D \omega)_{\mu_1 \dots \mu_p \lambda} = 0 \, ,
\end{equation}
and under local Weyl rescalings according to (\ref{one},\ref{two})
\begin{eqnarray}
\de_\si (\D \omega)_{\mu_1 \dots \mu_p \lambda} \!\! &=& \!\!
\half (d-2p-2) \, \si \, (\D \omega)_{\mu_1 \dots \mu_p \lambda} \nonumber \\
&& \!\!\!\! {} + \half (d-4) \bigg ( \pr_\lambda \si\, \omega_{\mu_1 \dots \mu_p}
- \pr_{(\mu_1} \si \,\omega_{\mu_2 \dots \mu_p)\lambda} \nonumber \\
&& \qquad \quad {} - \frac{p-1}{d+p-3} \Big (
g_{\lambda(\mu_1}
\omega_{\mu_2 \dots \mu_p)\rho } - g_{(\mu_1\mu_2}
\omega_{\mu_3 \dots \mu_p)\lambda\rho} \Big ) \pr^\rho \si \bigg ) \, .
\end{eqnarray}
Clearly when $d=4$ $\D$ is a first order 
conformally covariant differential operator.
Moreover from the definition (\ref{defD})
\begin{eqnarray}
\frac{p}{p+1}\, (\D \omega)^{\mu_1 \dots \mu_p \lambda}
(\D \omega)_{\mu_1 \dots \mu_p \lambda} \!\! &=& \!\! 
\nabla^\lambda \omega^{\mu_1\dots \mu_p}
\nabla_\lambda \omega_{\mu_1\dots \mu_p} 
- \nabla^\lambda \omega^{\mu_1\dots \mu_{p-1}\rho} 
\nabla_\rho \omega_{\mu_1\dots \mu_{p-1}\lambda} \nonumber \\
&& - \frac{p-1}{d+p-3}  \, \nabla_\rho
\omega^{\mu_1\dots \mu_{p-1}\rho} \nabla^\lambda
 \omega_{\mu_1\dots \mu_{p-1}\lambda} \, .
\label{exp}
\end{eqnarray}
By discarding total derivatives we may write
\begin{equation} \!\!
\nabla^\lambda \omega^{\mu_1\dots \mu_{p-1}\rho}
\nabla_\rho \omega_{\mu_1\dots \mu_{p-1}\lambda} \to
\nabla_\rho \omega^{\mu_1\dots \mu_{p-1}\rho} \nabla^\lambda
 \omega_{\mu_1\dots \mu_{p-1}\lambda}
- \omega^{\mu_1\dots \mu_{p-1}\rho} [ \nab_\lambda, \nab_\rho ] 
\omega_{\mu_1\dots \mu_{p-1}}{}^{\! \lambda} \, ,
\end{equation}
where, using the definition of the Weyl tensor in terms of the
curvature in (\ref{Weyl}) and also (\ref{}),
\begin{eqnarray} \!\!\!\!\!\!
\omega^{\mu_1\dots \mu_{p-1}\rho} [ \nab_\lambda, \nab_\rho ]
\omega_{\mu_1\dots \mu_{p-1}}{}^{\! \lambda} \!\! &=& \!\!
- (p-1)\, C_{\lambda\ep\rho\eta}\, \omega^{\mu_1\dots \mu_{p-2}\lambda\rho}
\omega_{\mu_1\dots \mu_{p-2}}{}^{\! \ep\eta} \nonumber \\
&& \!\!\!\!\!\!\!\!\!\! {}
+ (d+2p-4) K_{\lambda\rho} \, \omega^{\mu_1\dots \mu_{p-1}\lambda}
\omega_{\mu_1\dots \mu_{p-1}}{}^{\!\rho} + J \,
\omega^{\mu_1\dots \mu_p} \omega_{\mu_1\dots \mu_p} \, .
\end{eqnarray}
With these results, if we choose for the parameter $A$ in (\ref{S2})
$A=-(p-1)$, we may obtain an alternative expression for the total action
given by (\ref{S}),
\begin{eqnarray}
S[g,\omega] \!\!& =& \!\! \frac{1}{2p!} \int \! \d^d x \, \sqrt g \, \bigg [ 
\, \frac{p}{p+1}\, (\D \omega)^{\mu_1 \dots \mu_p \lambda}
(\D \omega)_{\mu_1 \dots \mu_p \lambda} \nonumber \\
&& \qquad\qquad {}
+ \frac{(d-4)(d-2)}{(d+2p-2)(d+p-3)} \, \nabla_\rho
\omega^{\mu_1\dots \mu_{p-1}\rho} \nabla^\lambda
 \omega_{\mu_1\dots \mu_{p-1}\lambda} \nonumber \\
&& \qquad\qquad {}
- (d-4) \Big ( K_{\lambda\rho}\, \omega^{\mu_1\dots \mu_{p-1}\lambda}
\omega_{\mu_1\dots \mu_{p-1}}{}^{\!\rho}- \half  J \,
\omega^{\mu_1\dots \mu_p} \omega_{\mu_1\dots \mu_p} \Big ) \bigg ] \, .
\label{SD}
\end{eqnarray}
This result demonstrates the importance of $d=4$, in this case only the first
term quadratic in operator $\D$ is present, which is in accord with the
results of Branson \cite{Branson}. The above formula (\ref{SD}), along  with
(\ref{exp}), coincides
with that given by O'Raifeartaigh {\it et al} \cite{Raifertaigh} who
required the absence of curvature dependent terms (although the motivation
for such a condition is not clear). If $p=1$ and $d=4$ (\ref{SD}) is manifestly
just the standard expression for conformally invariant Maxwell theory. On flat
space if, for some scalar $\rho$,
$\omega_{\mu_1 \dots \mu_p} = \pr_{\mu_1} \dots \pr_{\mu_p}\rho - {\rm traces}$
then $(\D \omega)_{\mu_1 \dots \mu_p \lambda} = 0$ which explains the absence of
an inverse in the flat space limit when $d=4$. Of course if $p=1$ this is just
a reflection of the usual gauge invariance of Maxwell's equations.

\end{document}